# Gauging Structural Aspects of ZnO nano-Crystal Growth ThroughX-ray Diffraction Studies and PAC

Bichitra Nandi Ganguly, SreetamaDutta, Soma Roy
Saha Institute of Nuclear Physics, Kolkata-700064, India

and

Jens Röder[1,2], Karl Johnston[1,3], ISOLDE-Collaboration[1]
[1.]Physics Department, ISOLDE/CERN, Geneva, Switzerland
[2.]Physical Chemistry, RWTH-Aachen, Aachen, Germany
[3.]Experimental Physics, University of the Saarland, Saarbrücken, Germany

*Corresponding author : Bichitra Nandi Ganguly*
*Saha Institute of Nuclear Physics, KOLKATA, INDIA*
*Fax :+ 91 33 23374637*
*Email : bichitra.ganguly@saha.ac.in*

# ABSTRACT

The structural characterization of sol-gel based nano crystalline ZnO material is being reported as we observe several previously unreported structural aspects following a sequence of annealing stages. As-grown samples were characterised by Fourier Transform Infrared Spectroscopy (FTIR). Chemical purity of the nano-grains and their crystallinity has been monitored by energy dispersive X-ray (EDAX) analysis and transmission electron microscopy (TEM), while the unusual changes in nano-crystal growth structure have been studied by X-ray diffraction method. In addition, such samples have been studied by using perturbed angular correlation (PAC) technique with the short-lived radioactive probe $^{111m}$Cd. Changes in the local electronic environment following sintering of the nano-crystalline grains have been observed by this method.

**Keywords :** ZnO nano crystal growth, X-ray diffraction, perturbed angular correlation (PAC), radio-active ion beam, hyperfine interaction.

## I. Introduction:

ZnO is one of the most versatile wide band gap (3.37eV) semiconductor materials known for its multifarious applications in technology [1,2]. The aqueous precursor–derived ZnO material is a promising alternative to organic semiconductors and amorphous silicon materials in applications such as transparent thin film transistors at low temperatures [3]. Also, the properties of ZnO transistors depend on its chemical preparation (such as extraction of the precipitate from acidic or basic solutions), drying and sintering processes and on the traces of impurities imparted to the system [4,5]. Chemical structure analysis suggests that the hydrated zinc cation, the pH of the medium and ligand type play a critical role in aqueous precursor based ZnO-nano crystalline material. These can affect any device related characteristics such as field effect mobility, drift velocity and switching ratios [6].

Also, ZnO is a bio-friendly oxide semiconductor and an inexpensive luminescent material. It is expected to have a wide range of applications in room temperature ultraviolet (UV) lasing [7], biosensors [8], bioimaging [9], drug delivery [10], piezoelectric transducers [11] and other usages as doped- ceramic compounds [12]. For all such purposes the growth of ZnO-nano scale material through a chemical route is a necessity. M**ost existing preparation techniques rely strongly on ZnO grain manipulation processes while drying and sintering the hydrated ZnO precipitate. Also, the defect structure in the evolution process of nano material is vital as there lies strong evidence that defects have a role to play in ferromagnetic order in such materials and the ferromagnetism coupling may be mediated by carriers**[13].

A systematic study therefore has been carried out to examine various physical aspects of chemically grown nano-crystal ZnO. Firstly, the chemical speciation and purity of the as-grown material, along with grain growth and crystallinity are checked using X-ray diffraction. Electronic microenvironment of the material is also probed using the PAC technique.

It is thus important to mention that a suitable radioactive dopant ion serving as PAC probe (for example :$^{111m}Cd^{+2}$, by exposure of the ZnO samples to radioactive ion beam at ISOLDE , CERN ) becomes a necessity. Further, facilitating the diffusion of the dopant ion into ZnO grains is followed by gradual and controlled annealing steps. An improvement of grain size consistency by controlling the formation of ultrafine grains and prevention of abnormal grain growth could yield a better result for fabricating nano-material for a device related purpose. Although we find ZnO nano-material has been studied in various ways, but systematic monitoring of the structural distribution in the presence of a dopant[5] at trace level in the host medium along with sintering processes are sparse. In this work, we have purportedly used $^{111m}Cd$ such that its direct route to stable state will not alter any other characteristics of the medium except the electric field gradient (EFG) tensor, which is the observable parameter from PAC measurement. Such a parameter is strongly dependent on the distribution of the electronic charge density around the site where the radioactive probe atom has been hosted in the lattice.

In this article, we develop a meaningful comparison of the changes that are observed for nano-crystal growth starting from the as-grown wet chemical stage through the sequential annealing steps, by X-ray powder diffraction pattern and PAC measurements.

## II. Experimental :

### A. Chemical Synthesis of Pure Zinc Oxide (ZnO)

ZnO nanoparticles were prepared by the sol-gel technique from zinc acetate ($Zn(CH_3COO)_2 \cdot 2H_2O$, extra pure AR grade material, from SRL, India). The desired weight of zinc acetate was dissolved in triple distilled water (TDW) and a (1:1/vol) ammonia solution (Merck India) was added to this solution drop by drop, maintaining a pH~7.5. Initially the zinc is precipitated as zinc hydroxide. After centrifugation, the precipitate was collected and re-dispersed into TDW to remove excess ions. Finally, the precipitate was recollected and dried at 100°C for 12 hours and sintered at different temperatures (200, 600 and 1000°C for 30 minutes) to evolve ZnO nano-crystallites. These ZnO nano-grains were characterized by Fourier transmission infrared (FT-IR) spectroscopy [14] (as pellets in KBr, without moisture) using a Perkin Elmer FTIR system, Spectrum 100) in the range of 400-4500 $cm^{-1}$, with a resolution of 0.4 $cm^{-1}$. The results are shown in Figure1 and Table I

### B Electron microscopic studies:

For the elemental purity check of the ZnO samples, energy dispersive X-ray analysis (EDAX/SEM) and scanning microscopic analysis of the grain surface was performed using Quanta FEG-200, FEi Company USA. TEM examination was performed with (Tecnai S-twin, FEI) using an accelerating voltage of 200 kV, having a resolution of ~ 1 Å. The results of such analysis are shown in Figure 2.

### C. Characterization with X-ray Powder Diffraction Measurements:

The phase structures of the samples were identified by X-ray diffraction, Rigaku TTRAX3 diffractometer with CuK$\alpha_{1,2}$ radiation($\lambda$ = 1.541 Å and 1.5444 Å) has been used. The data have been collected in the range (2θ) 10°–100° with a step size of 0.02°. Si has been used as an external standard to deconvolute the contribution of instrumental broadening [15]. The measured XRD pattern is shown in Figure 3.

Nano crystalline material usually suffers from structural strain as the grain interior is relatively defect free while the grain boundary consists of high-density defect clusters [16]. The size of ZnO nano grains and the corresponding strain in the lattice have been estimated by constructing a Williamson–Hall (W–H) plot, as shown by equation (1) taking different Bragg peaks [17] into consideration.

$$\beta \cos\theta = \frac{K\lambda}{D_{hkl}} + 4\,\epsilon\,\sin\theta \qquad (1)$$

where $\varepsilon$ is the micro strain parameter using the W-H plots using (equation 1), $\theta$ is in radians, $\beta$ = line broadening, $D_{hkl}$ = crystallite size and $K$ is the shape factor along with the wave length $\lambda$, the results are shown in Figure 4.

An estimation of the size of ZnO-nano grains and lattice parameters have been made by using the FullProf program [17] as shown in Table II. In addition, Rietveld refinement analysis was performed with GSAS [18], calibrating the instrumental parameters with a standard Si measurement. During the fits, instrumental parameters were kept constant. The results are shown in Figure 5.

From Thompson-Cox-Hastings pseudo-Voigt function (TCH) of the Lorentzian and Gaussian term, size and strain parameters can be retrieved directly from GSAS profile parameters [18]: using a Lorentzian TCH term with parameters GX and GY:

$$\Gamma_L = \frac{GX}{\cos\theta} + GY\tan\theta \qquad (2)$$

For the Williamson-Hall analysis, the integral breath (β) holds with Dv as the volume weighted average, $\varepsilon_{str}$ the strain and λ as the wavelength:

$$\beta_{obs} - \beta_{inst} = \frac{\lambda}{Dv\,\cos\theta} + 4\,\varepsilon_{str}\,\tan\theta \qquad (3)$$

A correction of the FWHM to integral breath and correcting GSAS centidegrees, size and strain can be calculated by:

$$Dv = \frac{36000\,\lambda}{\pi^2\,GX} \tag{4}$$

$$\epsilon_{str} = \frac{\pi^2(GY - GY_{inst})}{144000} \tag{5}$$

Gaussian TCH term with parameters GU and GP (GV and GP can be considered as instrumental constants):

$$\Gamma_G = GU\tan^2\theta + GV\tan\theta + GW + \frac{GP}{\cos\theta} \tag{6}$$

Williamson-Hall analysis:

$$\beta^2{}_{obs} - \beta^2{}_{inst} = \frac{\lambda^2}{(Dv)^2\,\cos^2\theta} + 16\,\epsilon^2\tan^2\theta \tag{7}$$

Applying the same corrections as for the Lorentzian part results in:

$$Dv = \frac{18000\,\lambda}{\sqrt{2\pi^3 GP}} \tag{8}$$

$$\epsilon_{str} = \frac{\sqrt{2\pi^3(GU - GU_{inst})}}{72000} \tag{9}$$

The volume weighted average $D_v$ can be converted into the diameter of spheres with identical size: d=4/3 $D_v$. (see Table II and IV)

**D. Local Structure Investigation with Perturbed Angular Correlation Using Radioactive Ion Beam of $^{111m}$Cd:**

The time differential perturbed angular correlation technique is based on the modulation of angular correlation of the successive radiations emitted during a nuclear decay cascade due to hyperfine interactions between the electromagnetic moments of the intermediate nuclear level with its immediate neighboring electronic environment. Suitable isotopes for standard PAC have an intermediate level (sensitive level) which has a half-life between 10 to 1000 ns (see Figure 6), while the half-life of the parent isotope is sufficiently long lived to provide a measurement. $^{111}$In with a 2.8 days half-life and simple cascade is one of the most straight forward and widely-used PAC probes. However, so-called *after effects*, resulting from the change of electronic charge state from $^{111}$In$^{3+}$ to $^{111}$Cd$^{2+}$, can be problematic especially in semiconductors [19]. In order to minimize these effects, the $^{111m}$Cd isotope is more suitable as it does not undergo any such transmutation. However, its short half life of 48 minutes restricts its availability and requires a facility such as ISOLDE at CERN.

A PAC machine consists of a setup of usually four or six detectors which measure the time difference between the $\gamma_1$- and $\gamma_2$-ray as shown in Figure 6. from which one obtains the radioactive decay curve of the intermediate level of $^{111m}$Cd with ~84.5 ns. Due to the anisotropy of the emitted radiation, ripples on the decay curve become visible when electric quadrupole or/and magnetic dipole interaction, e.g. electric field gradient (EFG) or local magnetic fields in magnetic materials, are non-zero as shown in Figure 6 (see in the middle).
The counting rate ratio, R(t), is formed from the coincidence data by:

$$R(t) = 2 \frac{N^{180°}(t) - N^{90°}(t)}{N^{180°}(t) + 2N^{90°}(t)} \tag{10}$$

The counting rate ratio can be described in practical use by [30, 31]:

$$R(t) \approx A_{22}^{eff} \sum_i f_i G_{22}^i(t) \tag{11}$$

With *A* as the effective anisotropy, *f* the fraction per site *i* and $G_{22}$ the perturbation factor. For quadrupole interaction with including the finite time resolution here in the last term, the following term was used for fitting the data:

$$G_{22}^i(t) = s_{20}^i(\eta_i) + \sum_{j=1}^{3} s_{2j}^i(\eta_i) \cos(\omega_j^i t) e^{-\frac{\omega_j^i \delta_i t}{v_{Q_i}}} e^{-\frac{(\omega_j^i \tau_r)^2}{16 \ln 2}} \tag{12}$$

Here, $\nu_Q$ as the quadrupole interaction constant and δ its distribution, ω the quadrupole circular frequency and τ the instrumental time resolution. More detailed description of the PAC method can be found at [20, 21, 22, 23].

Perturbed Angular Correlation (PAC) spectroscopy was performed in order to study the local structure in polycrystalline nano-ZnO (already mentioned). $^{111m}$Cd as PAC probe was implanted into cold pressed pellets of the previously prepared material. Room temperature implantations were performed at ISOLDE/CERN [24] at 30 keV. $^{111m}$Cd ions were produced following irradiation of a molten Sn target by 1.4 GeV protons, which were subsequently plasma ionized and mass separated . A typical irradiation dose was about $3 \times 10^{13}$ ions per sample.

PAC measurements were performed at room temperature (RT) using two Digital Time Differential PAC machines (DTDPAC) [25, 26] with four BaF$_2$ detectors resulting in 12 single spectra of 90° and 180° per measurement. All the detected γ-rays were saved as time and energy values on the DTDPAC machine's hard disks. Due to the short half-life of the probe, energy windows shifted slightly with the rather fast decrease of the activity. Therefore the recorded data were reprocessed by separating the data into several parts, determined for each part the optimal energy windows and merged the results of all parts finally together. The data were analyzed with XFIT program using the XPAC for XFIT program [27].

## III. Result and Discussion :

### A. FT-IR and EDAX Study:

ZnO samples prepared by aqueous sol-gel technique were dried under vacuum at 100 °C over night, its purity and the characteristic FT-IR frequencies were checked as shown in Figure 1. and Table I. These data represent the as-prepared ZnO material. Some residual acetate groups (C=O) could be detected. Presence of moisture is also detected, this could be dependent on the ambient condition.  This was considered as the precursor material for ZnO, which was sintered later to proceed for evolution of pure and dried material of ZnO. The elemental purity of ZnO was checked with the help of EDAX spectrum as shown in Figure 2. From TEM results, the fringe structure of ZnO material depicted crystallinity of the material (as shown by the arrow). The mean size as estimated from the TEM image has been about ~ 20 nm and clearly indicates that the initially annealed ZnO nanoparticles are crystalline with a wurtzite structure. No other impurities were observed.

**B. X-ray Diffraction (XRD) Study :**

The as-dried precipitate of ZnO material (at 100 ºC) can be considered as a precursor of the ZnO material without the typical ZnO phase, but mainly zinc acetate phase and an unknown phase. Further annealing of this sample stage by stage from (200 to 1000ºC) really stage by stage or 200, 600 and 1000 ºC, pure ZnO structure was produced, no characteristic peaks from intermediates such as $Zn(OH)_2$ could be detected in the samples. XRD results shown in Figure 3 give us the characteristic diffraction pattern of the crystallites under the particular configuration, through the Bragg angles. The appearance of characteristic diffraction peaks for a pure ZnO sample corresponding to (1 0 0), (0 0 2), (1 0 1), (1 0 2), (1 1 0), (1 0 3) and (1 1 2) planes is in good agreement with the standard XRD peaks of crystalline bulk ZnO with hexagonal wurtzite structure [JCPDS card No. 36–1451, a = 3.2501 Å , c = 5.2071 Å, space group: P63mc (1 8 6)], The gradual changes of the FWHM of characteristic peak (002) is shown as an inset to show the refinement of evolved structure, at 1000 ºC, the best result is obvious from the spectrum.

The size and strain analysis of the grains at different temperatures have been obtained following FullProf and W-H plot analysis and are shown in Table II and Figure 4. It is found that as the sintering temperature gradually increases, we obtain an increase in the crystal growth size and the lattice strain is found to be a maximum at 600ºC. Lattice strain generally arises due to vacancies, crystal imperfection, dislocations sinter stress, stacking faults indicating that ZnO nano-grain samples are undergoing structural changes under the sintering conditions.

Additionally, Rietveld refinement analysis was performed using GSAS and EXPGUI [28, 29]. For this, the instrumental parameters were determined by using a standard Si sample. (Fitted data according to GSAS labeling [28] were LX, LY, shft,ptec, lattice parameters, fractional coordinates and Uiso as well as spherical harmonics, preferred orientation with 6$^{th}$ order.) Lattice parameters are shown at different sintering temperatures in Table III and the unit cell parameters after sintering at 1000°C in Table IV.

While fitting these nano crystallite parameters, an unusual peak shape was observed which could not be fully adjusted by keeping the instrumental parameters fixed. This may indicate an irregular structure or larger variation in crystallite size. The fit could be improved by adding a Gaussian component indicated by parameters GU and GP, see Figure 5 for the sample sintered at 1000°C(in general, the Lorentzian term of the Thompson-Cox-Hastings pseudo-Voigt function fits sufficiently well-crystallized material with Bragg-Brentano X-ray powder diffractometers; Gaussian terms (GU, GP) are rarely used).

Another phenomenon of the samples are deviations in the expected intensities, which in these fits are compensated by using spherical harmonics preferred orientation in GSAS. However, for nano materials, producing textures during sample preparation in powder X-ray diffraction is less expected and can be explained by significant modifications in shapes as needles or platelets, the latter of these may form in ZnO samples, as is described by [30].

As shown in Table III, the crystalline size assumed for spherical particles of the same size increases steadily with increasing sintering temperature, while the lattice parameters stay almost unchanged. The fit of the sample sintered at 200°C could be significantly improved by adding a Gaussian term. The texture is found to be strongest at 600°C sintered sample and also represents the highest strained structure. The nano-crystallite size more or less corroborates from the fitting parameters by methods used, although the absolute value of strain may differ, but the trend is maintained from both the methods shown.

C. Perturbed γ-γ Angular Correlation Measurements:

In order to understand the changes in the local structure and the material properties induced due to subsequent annealing of ZnO, perturbed angular correlation spectroscopy (PAC) was performed with the probe $^{111m}$Cd and results are shown in Figures 7a, b, c.

The PAC experimental runs have been performed at room temperature, at first, without annealing ZnO sample and then after subsequent annealing of the ZnO samples at temperatures 600°C, and 1000°C respectively after implantation of $^{111m}$Cd for 30 minutes to release implantation defects. At room temperature the $^{111m}$Cd precursor ZnO sample shows no frequencies corresponding to standard ZnO, but a wide distribution of frequencies can be seen. The sample sintered at 600°C shows a known frequency from $\nu_Q$=30 MRad/s, (can be compared with earlier results [5]), with a wide distribution of about 4 MRad/s and η=0.47, while at 1000°C sintered material shows $\nu_Q$=31 MRad/s and η=0.11. This development indicates a steady growth of crystallinity, agreeing well with the X-ray diffraction growth of crystal size. The sample at 600°C indicates that many sites with co-exist with slightly different local structure. This may indicate a highly distorted environment and can be probably caused by higher shell quantity considering the core-shell model of small particles [5]. As expected, the asymmetry parameter (η) is larger, at this stage, in highly distorted environments.

As per the density functional theory calculations [5] of the lattice sites, EFG have been performed with good agreement of Cd impurity in to ZnO the measurements. This confirms well with the expectations for incorporation of $Cd^{2+}$ in ZnO as both the metals are from the same IIB group of elements, having similar crystal radii [31] with $Cd^{2+}$ in tetragonal coordination with $r_{Cd+2}$ = 0.97 Å and for $Zn^{2+}$ in equivalent environment with $r_{Zn+2}$=0.74 Å. As $Cd^{2+}$ is larger than $Zn^{2+}$, the local structure may be slightly distorted, which may cause a larger effect in highly distorted material, where restoring forces are smaller. The final sintering at high temperatures of 1000°C transforms the material in normal bulk ZnO.

The performed measurements suggest that the preparation of nano-ZnO with the sol-gel method provides different materials properties which are possibly scalable with sintering conditions.

## VI. Conclusion:

Sol-gel prepared and grown nano-ZnO showed interesting properties as observed with FT-IR spectroscopy, X-ray diffraction and perturbed angular correlation method. It shows different structural aspects of the nano-material which *hitherto* remained unrevealed by a systematic study.

Material sintered at 200°C already forms ZnO with small crystallite size between 20- 26 nm but continuously growing with tempering at higher temperature. X-ray diffraction studies indicate that the produced material, depending on sintering properties varies significantly its structural properties. Especially interesting is the fact, that for medium tempered nano-ZnO, the peak profile shapes are different than the regular crystalline material, which indicates significantly different properties.

Perturbed angular correlation using the $^{111m}$Cd probe confirmed the results with previous works and shows that the material at lower sintering temperature posses a higher order distorted local structure, which may be explained through the core-shell model.

Structurally different, evolving nano-ZnO crystalline grains could be further useful for future study of semi-conductor application oriented requirements like thin film transistors, opto-electonic devices and development of flexible electronics etc.

## V. Reference:

[1] Zhong Lin Wang, J.Phys, : Condens. Matter **16,** R829-R858(2004**).**

[2] Nomura K. Ota, H. Tagaki, T. Kamiya, M. Hirano, H**.** Hosono, Nature **432,** 488-492 **(** 2004**).**

[3] H. E.A Huitema,. G. H. Gelinck, J. B. P. H.van der Putten, K. E. Kuijk, C. M., E. Hart, P. T. Cantatore, , A. J. Herwig, J. M. van Breemen and D. M.de Leeuw, Nature , **414**, 599 (2001).

[4] Taehwan Jun, Yangho Jung, Keunkyu Song and Jooho Moon, Applied Materials and Interfaces, **3,** 774-781 **(**2011).

[5] L.Emiliano, E. L. Munoz , M. E. Munoz, M. R. Mercurio, L.F.D. Cordeiro, A.W. Pereira, M. Carbonari, Physica B. **407,** 3121-3124 (2012).

[6] F. Fleischhaker, V. Wloka, and I. Hennig, J. Mater. Chem. **20**, 6622 –6625 **(**2010).

[7] Z.K Tang,G. K. L. Wong, P. Yu *et al*,. Appl Phys Lett **72,** 3270-3272 (1998).

[8] Z. Zhao, X. Lei. Zhang, B. Wang, and H. Jiang, Sensors **10**, 1216-1231(2010)**.**

[9] K. Senthilkumar, O. Senthilkumar, Kazuki Yamauchi,Sato Moriyuki, Morito Shigekazu, Ohba Takuya, Nakamura Morihiko, Fujita Yasuhisa, Phys.Status Solidi B, **246,** 885-888 (2009).

[10] Y. Wang, L. Chen, Nanomedicine: Nanotechnology, Biology and Medicine, **7**, 385-402.(2011).

[11] D.C. Oertel, M. G.Bawendi, A.Arango, C.V. Bulovic, Appl Phys Lett., **87,** 213505-213507 (2005**).**

[12] D.R. Clarke, J Am Ceram Soc., **82,** 485-502 (1999).

[13] M. Khalid, M. Ziese, A. Setzer, P. Esquinazi, M Lorenz, H. Hochmuth, M. Grundmann, D.Spemann, T. Butz, G.Brauer, W.Anwand, G. Fischer, W. A Adeagbo, . W. Hergert, and **A.** Ernst Physical Review B. **80**, 035331 (5 pages)( 2009**)**

[14] L.J.Bellamy., Infrared Spectra of Complex Molecules, (Methuen,**1959** ) London.

[15] B.D.Cullity, S.R. Stock: *Elements of X-ray Diffraction* **(**Prentice-Hall, **2001)** Englewood Cliffs, New Jersey.

[16] Tichy Ungár, Géza Tamás , Jenő Gubicza, and R. J. Hellmig, Powder Diffr, **20**, 366-375(2005).

[17] G.K. Williamson, and W.H. Hall, Acta Metall. **1,** 22-31(1953**).**

18. P. Karen, P.M. Woodward, J. Solid State Chem. **141,** 78-88 (1998).
19. L. Dundon, Dissertation: R. D. Oregon State University, 1992 . Oregon, USA.
20. H. Frauenfelder, R.M. Steffen, Alpha-, Beta-, Gamma-Ray Spectroscopy, Ed: .Siegbahn, vol 2, ( North Holland, Amsterdam, 1965**)** Ch XIX, Page 997.
21. A Weidinger, M. Deicher, Hyperfine Interactions **10** , 717-720 **(**1981).
22. J. Röder, K.D. Becker , Methods in Physical Chemistry, Ed: R.Schäfer, P. C. Schmidt: **(**Wiley - VCH, Weinheim 2012).
23. E. N. Kaufmann, R. J Vianden, Reviews of Modern Physics **51,** 161-214 (1979)
24. E. Kugler, Hyperfine Interactions **129**, 23–42**(**2000).
25. C Herden, J. Röder, J.A. Gardner, K.D. Becker, Nuclear Instruments and Methods in Physics Research Section A. **594**, 155-161 (2008).
26. J. Röder, C. Herden, J. Gardner, A. Becker, K. D. Uhrmacher, M. H. Hofsäss, Hyperfine Interactions **181**, 131-139 (2008).
27. Röder, J. PAC Spektroskopie an Ruddlesden-Popper Phasenund methodische Entwicklungen zur Digitalen PAC-Spektroskopie, **(**Sierke Verlag **2009)**.
28. A.C. Larson and R.B. Von Dreele, General Structure Analysis System (GSAS), (Los Alamos National Laboratory Report LAUR 2000**),** p 86-748.
29. B. H. Toby, J. Appl. Cryst.**, 34,** 210-213 **(2001).**
30. G. Han, A. Shibukawa, M. Okada, Y. Neo, T. Aoki, H. Mimura, J. Vac. Sci. Technol. B **28** , C2C16 (2010).
31. R. D. Shannon, Acta Crystallographica A **32**, 751-767(1976).

## VI. Captions to the Figures :

Figure 1: FT-IR spectrum representing the characteristic frequencies of ZnO material synthesized from aqueous route and dried in vacuum oven at 100 ⁰C for 12 hours and then dried to 200 ⁰C.

Figure 2: EDAX data: a purity check of the ZnO sample as synthesized through sol-gel chemical route, inset shows the fringed structure of the crystalline sample from TEM picture.

Figure 3: X-ray diffraction pattern (raw data as obtained) ZnO-nano crystalline grains at different sintering temperatures.

Figure 4: ZnO nano grain size from W-H plot by X-ray analysis at different sintering temperatures.

Figure 5 : X-ray diffraction spectrum with Rietveld refined pattern of ZnO annealed at 1000⁰C , showing the Bragg peaks positions, the data points(+++), the fitted curve (___) ,and the difference curve(__).

Figure 6 : Decay scheme of $^{111m}Cd_{48}$, IT , **T$_{1/2}$=48.54m** , isomeric level with 5/2$^+$ spin where EFG tensor interacts with nuclear quadrupole moment Q, an illustration of the typical PAC measurement.

Figure 7 (a) : PAC measurements after doping $^{111m}$Cd at R.T., without annealing, showing the experimental R(t) spectrum and the hyperfine interactions;
(b) :PAC measurements after doping $^{111m}$Cd and after annealing to 600⁰C, showing the experimental R(t) spectrum and the hyperfine interactions;
(c) : ZnO nano crystalline material was doped with Cd$^{111m}$ in ISOLDE experiment and annealed at 1000 °C, measured in PAC set up . The experimental R(t) spectrum shows well defined hyperfine interaction and the Fourier transforms of the hyperfine parameters are also represented (top).

## VII. Tables

**Table : I** IR spectroscopic group frequencies[14] for the prepared pure ZnO nano crystalline sample, dried at 100⁰C and compared to that after drying to 200 ⁰C (refer to Figure 1.)

| Absorption wave number cm-$^1$ | functional Group frequencies |
|---|---|
| ~500 , 525 -560, | Zn—O stretching |
| ~1020 | H-O-H bending |
| 1430, 1570 | —C=O stretching |
| ~3500 and greater | —OH stretching |

**Table II**: Lattice parameters of ZnO-nano crystalline grains(hexagonal/wurtzite) by FullProf program:

| T [°C] | c [Å] | a [Å] | c/a |
|---|---|---|---|
| 200 | 4.93138 | 2.9874 | 1.65073 |
| 600 | 5.2015 | 3.20101 | 1.62496 |
| 1000 | 5.20657 | 3.18397 | 1.63524 |

**Table III** : Lattice parameters at different sintering temperatures as per the Rietveld refinement analysis using GSAS.

| T [°C] | a [Å] | c [Å] | Texture Index (GSAS) | Strain [%] | Size [nm] (diameter of spherical particles) |
|---|---|---|---|---|---|
| 200 | 3.2500 | 5.2070 | 1.078 | 0.0007 | 26 |
| 600 | 3.2496 | 5.2057 | 1.147 | 0.0890 | 46 |
| 1000 | 3.2499 | 5.2056 | 1.021 | 0.0222 | 76 |

**Table IV** : The unit cell parameters after sintering at 1000°C as per Rietveld refinement analysis

| Spacegroup | a [Å] | c [Å] | α, β [°] | γ [°] |
|---|---|---|---|---|
| P 63 m c | 3.2499 | 5.2056 | 90 | 120 |
| **Atom/Charge** | **x** | **y** | **z** | **100*Uiso** |
| $Zn^{+2}$ | 0.3333 | 0.6666 | 0 | 2.917 |
| $O^{-2}$ | 0.3333 | 0.6666 | 0.3759 | 3.517 |

**Table V:** Hyperfine interaction parameters of ZnO derived from the PAC in the ZnO samples showing profound effect on annealing temperature (refer to Figure 7.)

| T [°C] (annealing temperature) | $A_{22}$ | $\nu_Q$ [MRad/s] | $\delta$ [MRad/s] | $\eta$ |
|---|---|---|---|---|
| 600 | 0.056 | 30.58 | 4.145 | 0.463 |
| 1000 | 0.091 | 31.25 | 0.543 | 0.000 |

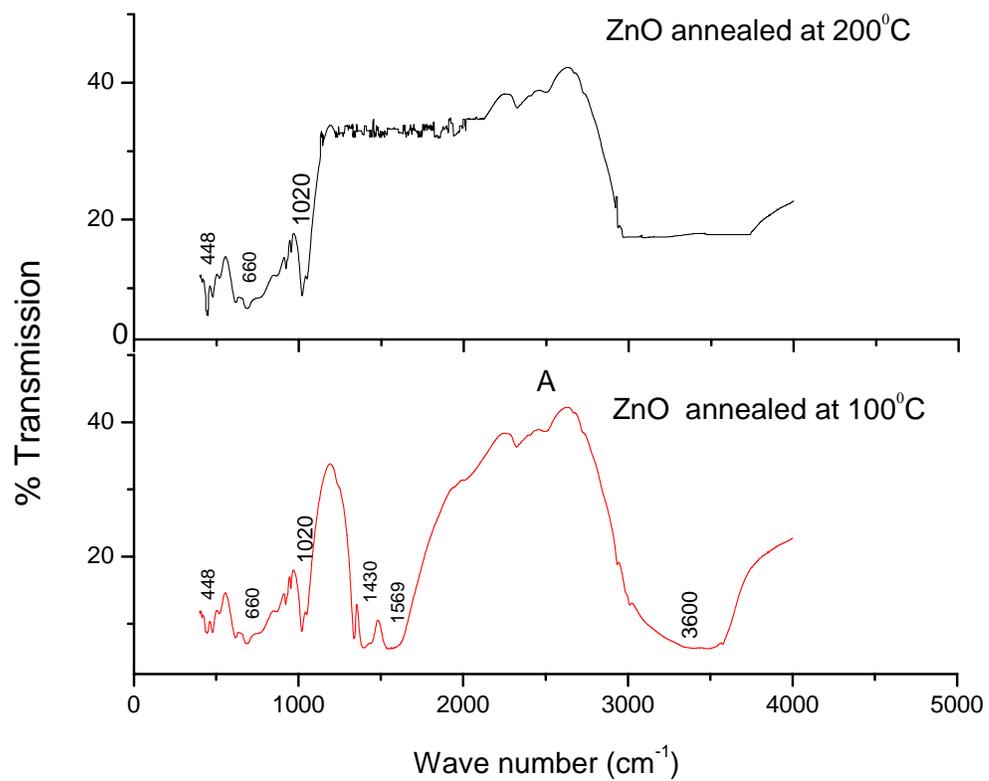

**Figure .1.**

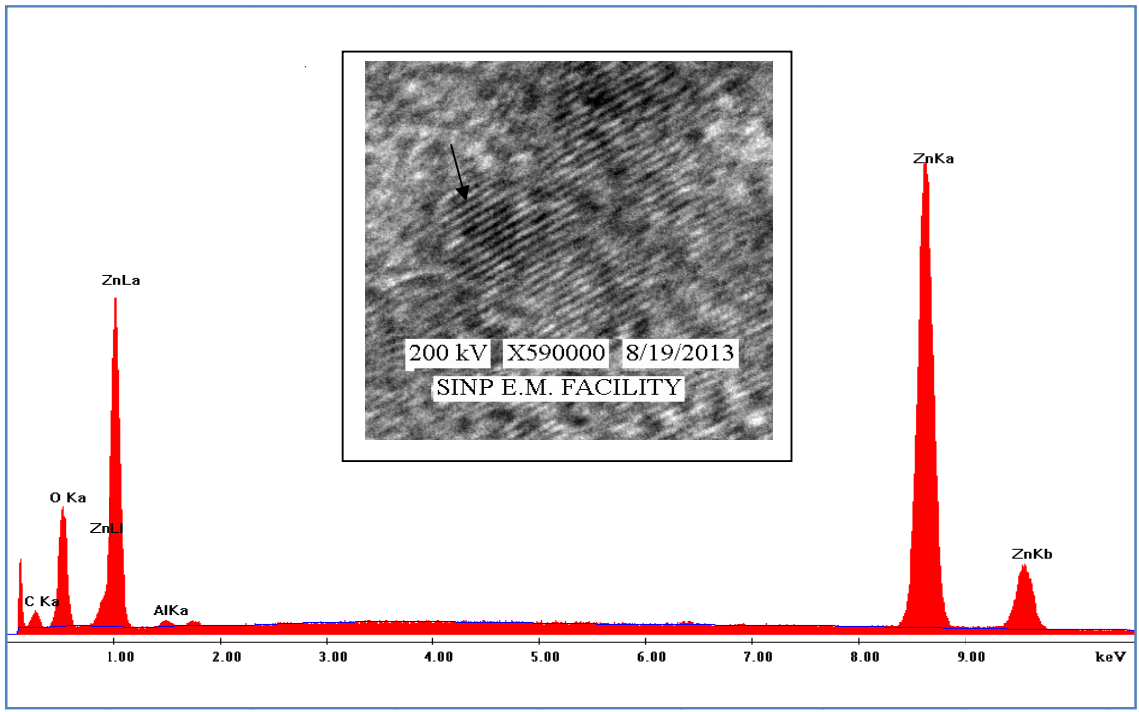

**Figure 2.**

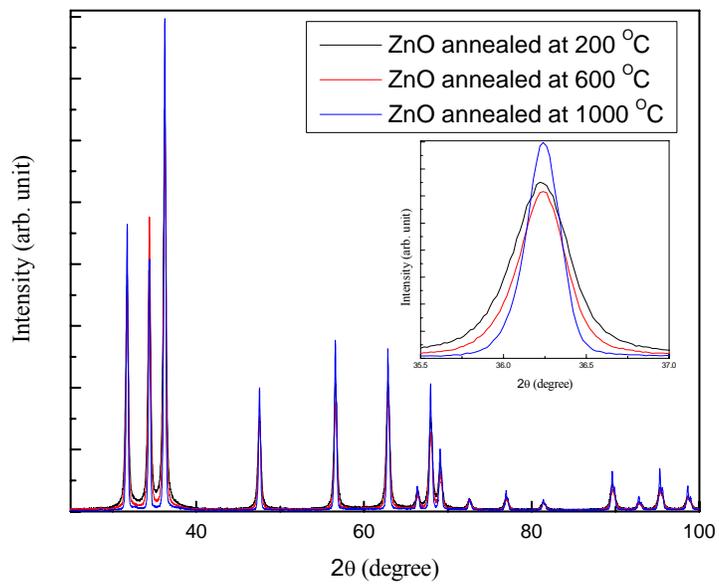

**Figure. 3**

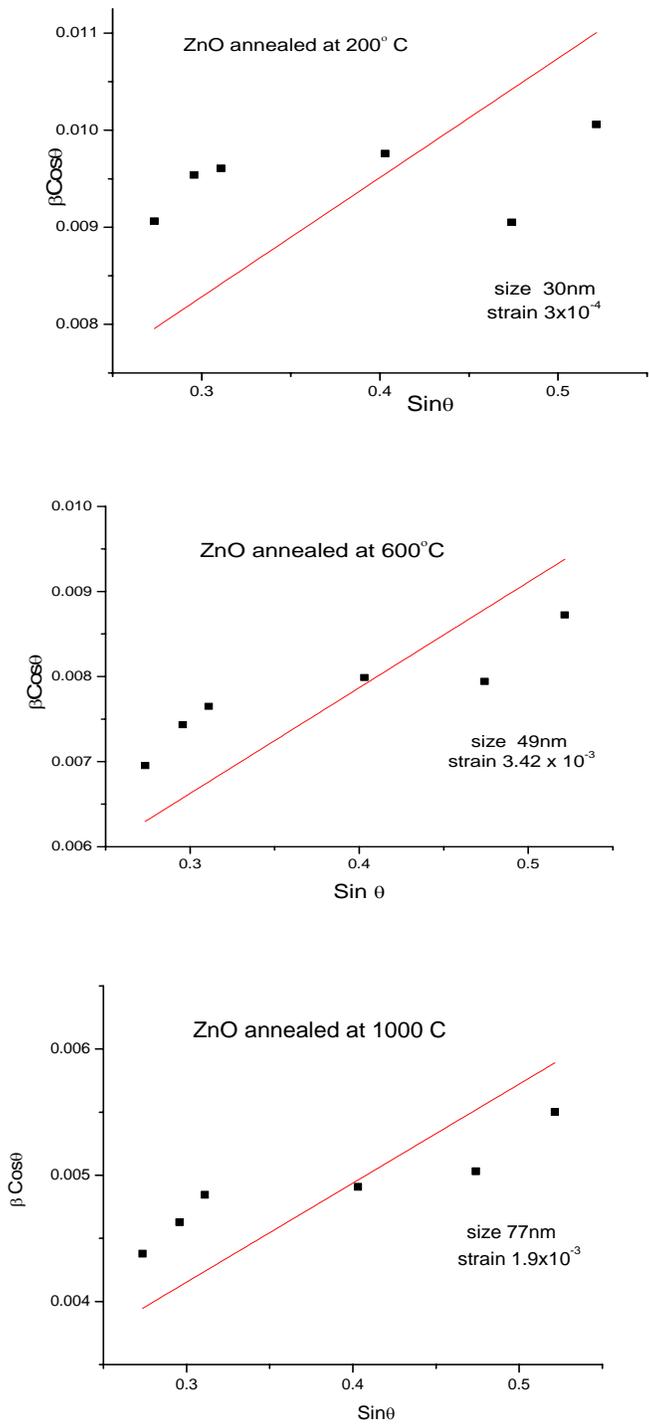

**Figure.4**

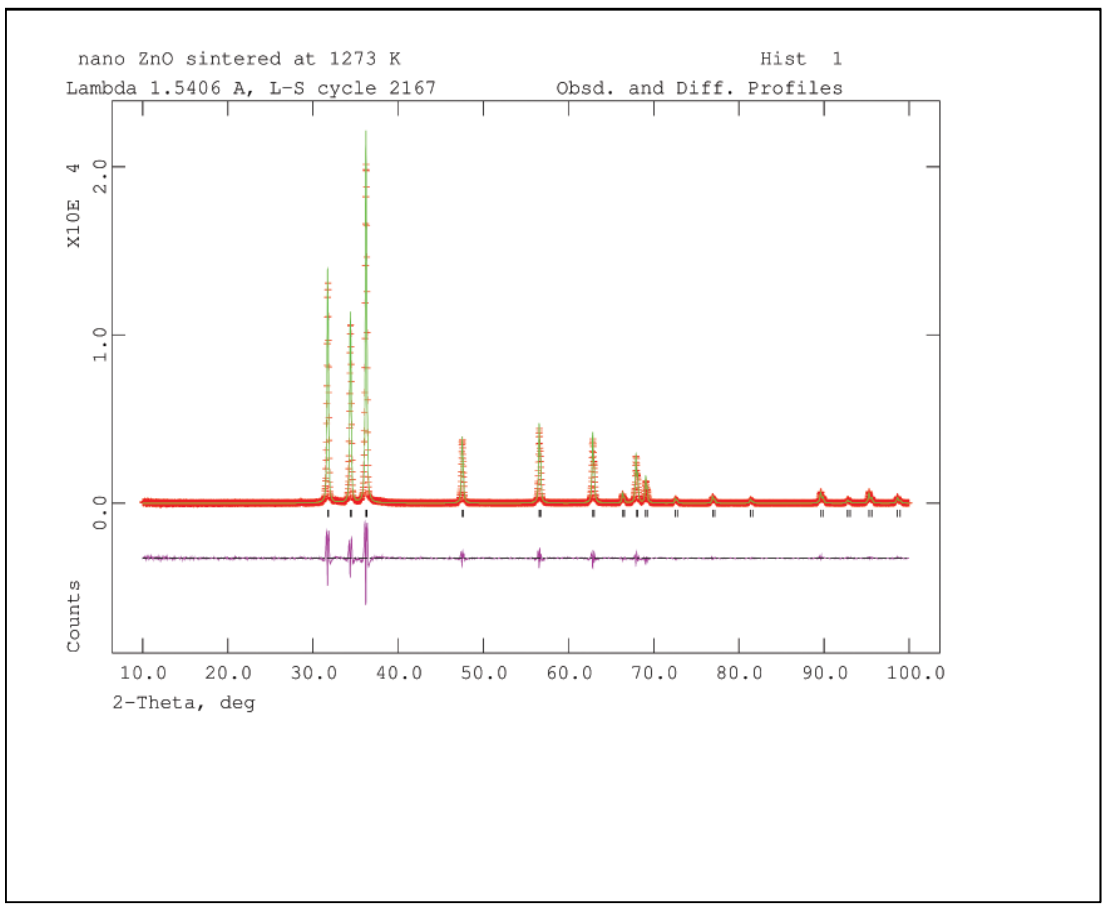

**Figure. 5**

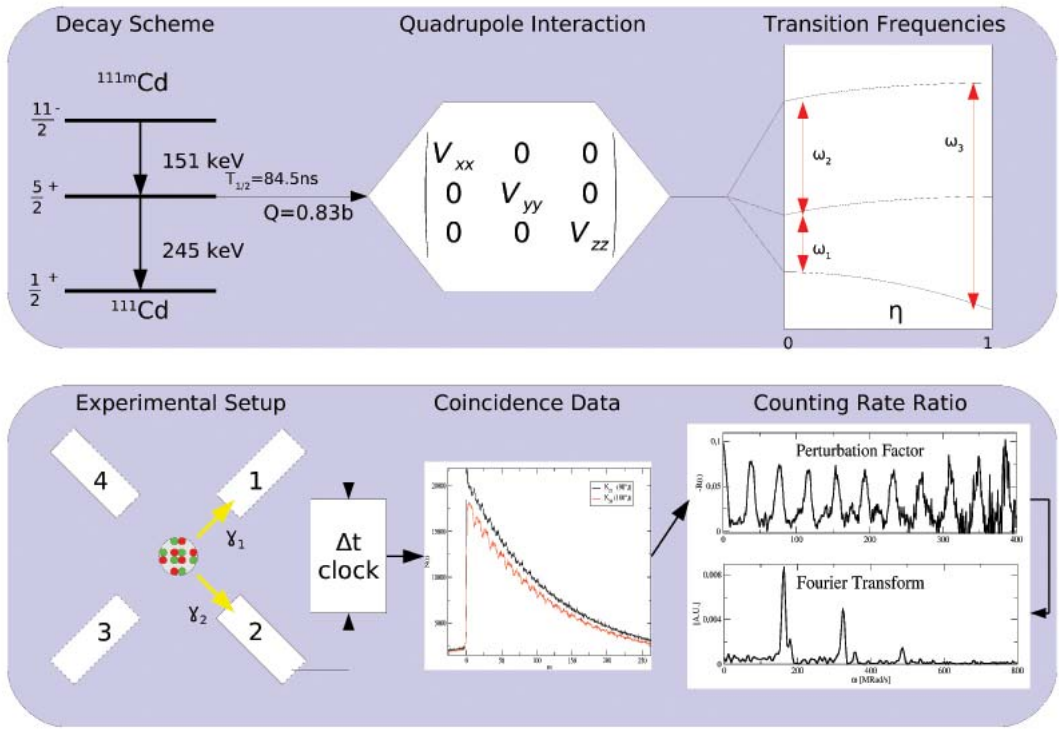

**Figure. 6**

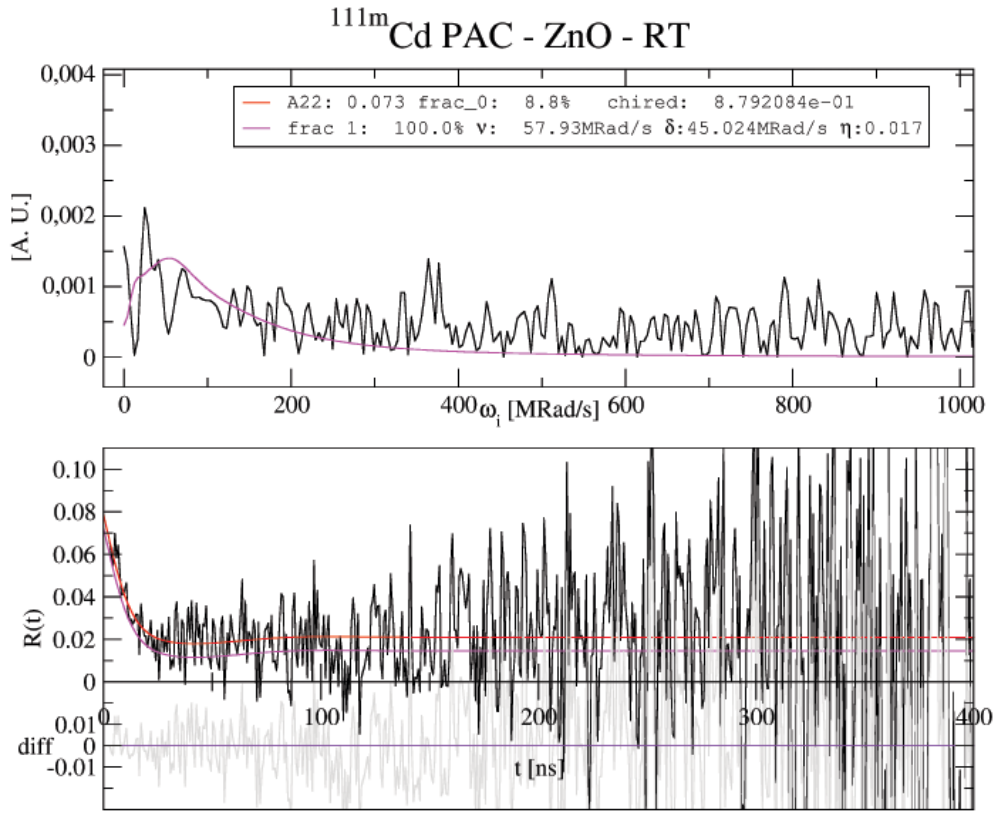

**Figure. 7a**

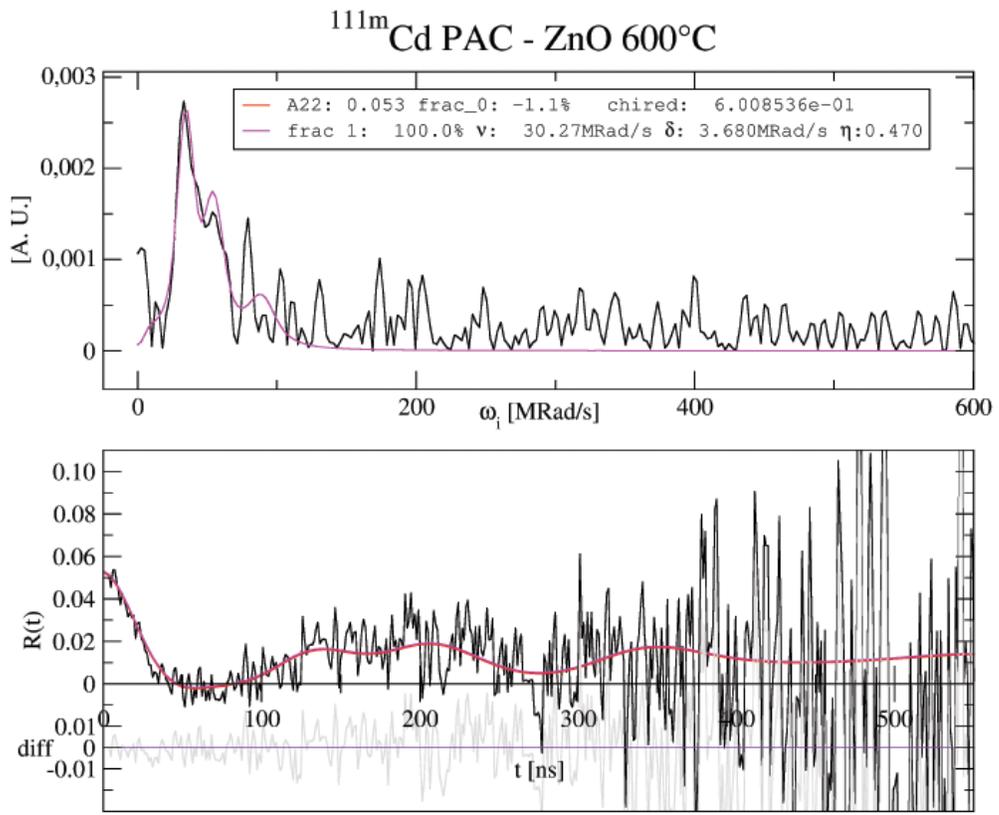

**Figure. 7b**

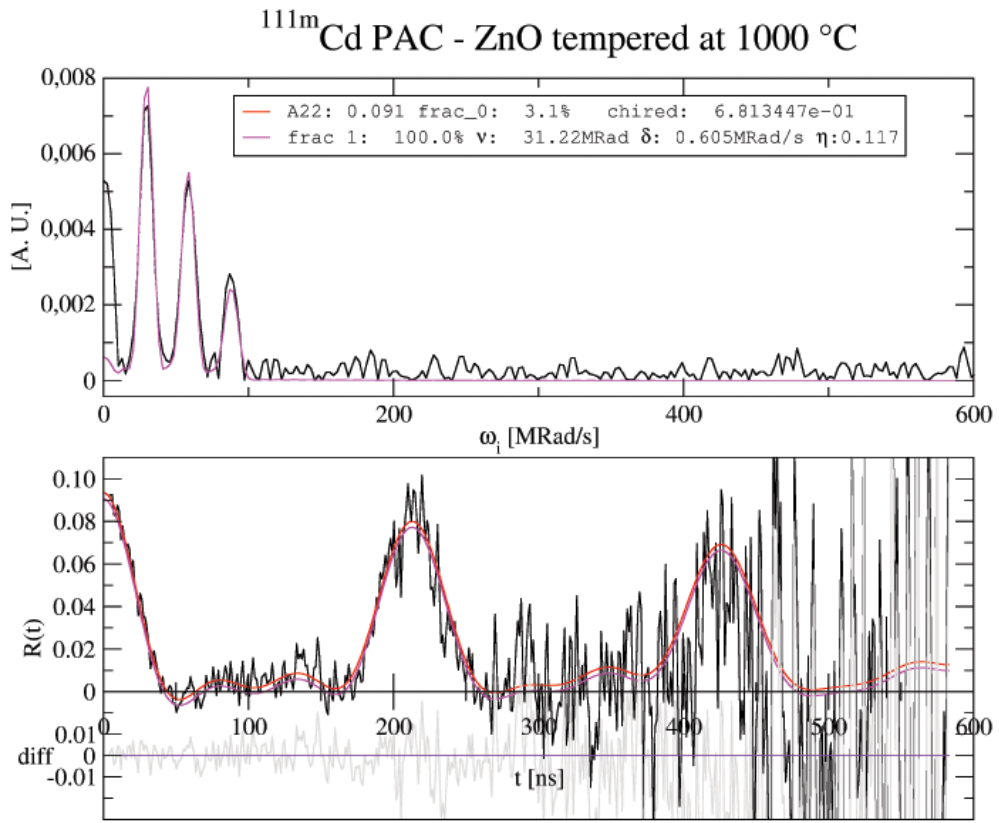

**Figure. 7c**